 \documentstyle[preprint,aps,tighten,prb]{revtex}

\newcommand{\be}{\begin{equation}}
\newcommand{\ee}{\end{equation}}
\newcommand{\bea}{\begin{eqnarray}}
\newcommand{\eea}{\end{eqnarray}}
 
\begin{document} 
\draft 

%%%%%%%%%%%%%%%%%%%%%%%%%%%%%%%%%%%%%%%%%%%%%%%%%%%%%%%%%%%%%%%%%%%%%%%%%%%%%%
\title{Universal Transverse Conductance between Quantum Hall Regions
and $(2+1)$D Bosonization}
%%%%%%%%%%%%%%%%%%%%%%%%%%%%%%%%%%%%%%%%%%%%%%%%%%%%%%%%%%%%%%%%%%%%%%%%%%%%%%

\author{Daniel G.\ Barci$^{a,b}$ and Luis\ E.\  Oxman$^c$} 
 
\address{a)  \it Department of Physics, University of Illinois at 
Urbana-Champaign, 1110
W. Green St., Urbana, IL  61801-3080, USA \\
{b) \it Departamento de F\'\i sica Te\' orica, 
Universidade do Estado do Rio de Janeiro, Rua S\~ao Francisco 
Xavier 524, 20550-013 Rio de Janeiro, RJ,Brazil
\footnote{Permanent Address \\
barci@dft.if.uerj.br \\
oxman@fis.puc-rio.br 
}}\\
{c) \it Departamento de F\'\i sica, Pont\'{\i}ficia Universidade
Cat\'olica do Rio de Janeiro,\\ 
Cx. Postal 38071, 22452-970, Rio de Janeiro, RJ, Brazil}}

\date{March 3, 2000} 
 
\maketitle 
%address: }
% \thanks{
%barci@dft.if.uerj.br,
%%%%%%%%%%%%%%%%%%%%%%%%%%%%%%%%%%%%%%%%%%%%%%%%%%%%%%%%%%%%%%%%%%%%%%% 
\begin{abstract}
%%%%%%%%%%%%%%%%%%%%%%%%%%%%%%%%%%%%%%%%%%%%%%%%%%%%%%%%%%%%%%%%%%%%%%%

Using bosonization techniques for $(2+1)$D systems, we show that the transverse
conductance for a system with general current interactions, when measured
between perfect Hall regions is not renormalized at low temperatures. Our
method extends two results we have recently obtained on low
dimensional fermionic systems: on the one hand,
the relationship between universality of Landauer conductance and
universality of bosonization 
rules for $(1+1)$D systems, and on the other hand, the universal
character of the bosonized 
topological current associated to a $(2+1)$D fermionic system with
current interactions. 

\end{abstract} 

%%%%%%%%%%%%%%%%%%%%%%%%%%%%%%%%%%%%%%%%%%%%%%%%%%%%%%%%%%%%%%%%%%%%%%%

{\bf Keywords:} Path integral bosonization, Universal Transport properties, 
Quantum Hall Effect.

{ \bf Preprint:} cond-mat/0001022

\pacs{PACS numbers: 11.10.-z , 03.65.Db, 11.25.Sq, 11.15 -q,11.15.Tk }

% 11.a0.-z Field Theory
% 11.25.Sq Nonperturbative techniques; string fields 
% 11.15 -q Gauge filed theory
% 11.40.Dw General Theory of currents
% 11.15.Tk Other nonperturbatives techniques
% 03.65.Db Functional Analitic Methods
% 71.45.Lr Charge-density-wave systems
% 72.10.Bg General formulation of transport theory
% 72.15.Nj Collective modes
% 72.15.-v Electronic conduction in metals and alloys
% 73.23.-b Mesoscopic systems
% 73.23.Ad Ballistic transport
% 73.20.Dx Electronic states in low-dimensional structures
% 73.40.Rw Metal-insulator-metal structures
% 05.60.-k Transport processes
% 11.30.Rd Chiral symmetries
% 11.40.Dw General theory of currents

%\begin{multicols}{2}
 
\narrowtext

%%%%%%%%%%%%%%%%%%%%%%%%%%%%%%%%%%%%%%%%%%%%%%%%%%%%%%%%%%%%%%%%%%%%%%%
\section{Introduction} 
%%%%%%%%%%%%%%%%%%%%%%%%%%%%%%%%%%%%%%%%%%%%%%%%%%%%%%%%%%%%%%%%%%%%%%%

Universal properties of interacting fermionic systems have always attracted
the attention of physicists. The reason is that a clean and exact behavior
for a system containing impurities and complicated particle interactions must
be associated to a strong constraint imposed by a simple physical principle.
Low dimensional condensed matter systems offer a variety of situations where
these universal phenomena occur.

For 1D systems such as quantum wires, recent experiments\cite{Tarucha95}
showed that, at low temperatures, the measured Landauer conductance is equal
to the quantum $e^2/h$ for each propagating channel. Soon it was understood
that, for a {\it finite} Luttinger liquid wire, the conductance is not
renormalized by the interaction in the
wire\cite{Maslov95,Ponomarenko95,Safi95}, since it is dominated by the
noninteracting electron gas in the leads. Moreover, 
a general relationship between universality in transport properties of 1D
systems and chiral symmetry has been recently proposed\cite{Alekseev98,OMK}. 

In Ref. \onlinecite{OMK}, following Maslov and Stone's
reasoning\cite{Maslov95}, one of us showed that perfect
conductance should also occur, at low temperatures, for a 1D
incommensurate charge density wave (CDW) system adiabatically connected
to Fermi liquid leads. This result 
agrees with that obtained in Ref.~\onlinecite{Rejaei96}, where transport of
charge in disordered mesoscopic CDW heterostructures was studied within the
Keldysh formalism. In contrast to the elaborate calculation of
Ref.~\onlinecite{Rejaei96}, we presented very simple physical arguments
based on the existence of a chiral (anomalous) symmetry for the system as a
whole (CDW plus leads) when the phase of the CDW order parameter is dynamic.

In Ref. \onlinecite{OMK}, we also stressed the important role played by
the finitness of the system, the adiabatic contacts to the reservoirs, and
the universal character of the bosonization rules, showing that a
general 1D structure of the 
Fermi-liquid/finite-system/Fermi-liquid type displays a perfect Landauer
conductance at low temperatures, provided the finite system presents an
(anomalous) chiral symmetry. The adiabaticity allows the extension of the
symmetry to the system as a whole in such a way that an anomalous chiral
current is always present when a bias voltage is applied; outside the
sample, this current is associated to the transport of free fermions. 

These general properties cause the charge transport through the system to be
dominated by the reservoirs, i.e., chiral symmetry is the physical principle
behind the universality of Landauer conductance in 1D systems. The natural
language we used to study these systems is bosonization, which maps the
initial $(1+1)$D fermionic system into a bosonic one, describing collective
excitations represented by a scalar field
\cite{Col-Man,BQ2,Naon,bosonizationbook}.

In connection with 2D systems, the understanding of the impurity
independence of the transverse conductance in the quantum Hall effect was
initiated by Prange \cite{Prange}, but it was after the work by
Laughlin\cite{Laughlin} and Halperin\cite{Halperin} that the underlying
mechanism for the universal character of the transverse conductance was
associated to the principle of gauge invariance. This is the accepted
explanation for the amazing degree of accuracy for the transverse
conductance, which is insensitive to such details as the sample's geometry
and the amount of impurities. A very interesting topological interpretation of
this fact can be found in Ref. \onlinecite{Thouless}.

The aim of this work is to show that bosonization is also a natural
language to describe universal transport properties for $2$D systems.

The generalization of the bosonization technique to higher
dimensions is recent.
In the context of condensed matter systems, the first
attemps to bosonize a Fermi-liquid in higher dimensions were presented in
Refs. \onlinecite{Luther} and \onlinecite{Haldane}. In Ref.
\onlinecite{CastroNeto}
the shape fluctuations of the Fermi surface were studied in a
very systematic way, and a detailed analysis of the Landau theory as a fixed
point of the renormalization group were presented.

In the context of quantum field theory, an
important activity on bosonization in higher dimensions was initiated
in the beginning of the nineties,
\cite{Marino,FF,Fidel,BQ1,BFO,BOS,c1,Banerjee}.
In particular, the bosonization of a massive $(2+1)$D Dirac field is 
achieved in terms of a gauge theory, where the Chern-Simons action plays a 
fundamental role, and the
fermionic current is mapped into the topological current $\epsilon^{\mu \nu
\rho}\partial_\nu A_\rho$. This bosonization, in contrast to the
abovementioned Fermi-liquid case, deals with parity breaking systems
(in $(2+1)$ dimensions, the mass term $M\bar{\psi}\psi$ breaks parity).
In this paper, this is the kind of systems we will be interested in, namely,
$2$D systems displaying the following properties:

\begin{itemize}
\item gauge invariance (charge conservation).

\item  a gap in the low lying charged excitations. 

\item Lorentz or Galilean invariance.

\item Parity or time reversal symmetry breaking.

\end{itemize}

Besides studying the $(2+1)$D fermionic case, which is the simplest one, we
will also study the physically relevant nonrelativistic case, where 
$2$D spinless fermions are submetted to an external magnetic field
$B$; here the gaps are provided by the Landau quantization.

It is worthwile stressing here that the two examples we will consider
have a quite different underlying physics. One of the consequences is 
that the particular values for the universal transport properties of these
systems will be different; however, in both cases, the proof of
universality will be similar as it relies on the general properties
shared by them. The role played by bosonization is to implement
all these properties in a very simple and compact way.

Esentially, we will extend two results we have recently
obtained for low dimensional fermionic systems. One of them is the
relationship between the universality of Landauer
conductance and the universality of the mapping between the fermionic
current and the bosonized topological current $\epsilon^{\mu
\nu}\partial_\nu \phi$, in $1$D systems\cite{OMK}.
The other one, is the universal character of the bosonized topological
current $\epsilon^{\mu \nu
\rho}\partial_\nu A_\rho$, for a class of $(2+1)$D 
systems \cite{BOS}; this will be related to the universality
of the transverse conductance, for a system with general current
interactions, when measured between ``perfect Hall regions''
(where the parity breaking parameter, $M$ or $B$, goes to infinity).
As a byproduct, we shall also see that the Aharonov and Casher
results\cite{Aharonov-Casher} for fermionic systems with spin, will
remain valid when interactions are included.

Thus, we will see that bosonization is a method that unifies universal
physical behaviors associated to systems with different dimensionality. While, 
in $1$D, bosonization is a simple way to display the anomalous 
properties of chiral symmetry, in $2$D, it is appropriate to display 
the gauge invariance of 
the effective fermionic action, that is, the physical
principle behind the universal character of transverse conductance. 
Then, it is no by chance that, in this framework, we
shall be able to derive the universal transverse transport. 

Although a closed expression for the bosonized action in higher
dimensions is still lacking, 
the gauge and topological structure of the bosonized theory, and the
universal character of the bosonized currents are the only properties 
we shall need to derive our results.
In analogy with the $1D$ case, where the Fermi-liquid in the reservoirs
imposes strong constraints on the Landauer conductance, we shall see that
perfect Hall regions will impose strong constraints on the
transverse conductance. As before, an adiabatic transition between the
interacting and the noninteracting regions will be needed. From a physical 
point of view, this condition corresponds to nondissipative contacts.

This paper is organized as follows: In section \ref{bos} we review and
compare the functional bosonization technique in $(1+1)$D and $(2+1)$D. 
In \S\ref{1D} we briefly review the relationship between the universality 
of the bosonization rules in $(1+1)$D and the universality of Landauer 
conductance in $1$D finite systems. Section \ref{2D} has the main results 
of this paper where we deduce the universal transverse conductance between 
``perfect Hall regions'', using the universal mappings between currents. 
In \S\ref{nonrel} we extend our results to the nonrelativistic case,
obtaining the universal properties of transverse currents, in  
the integer as well as in the fractional quantum Hall effect. 
Finally in \S\ref{conclusion} we discuss our results and give our 
conclusions. 

%%%%%%%%%%%%%%%%%%%%%%%%%%%%%%%%%%%%%%%%%%%%%%%%%%%%%%%%%%%%%%%%%%%%%%%%
\section{The bosonization technique }
%%%%%%%%%%%%%%%%%%%%%%%%%%%%%%%%%%%%%%%%%%%%%%%%%%%%%%%%%%%%%%%%%%%%%%%%
\label{bos}

In order to deal with the general bosonization structure in higher
dimensions, it is convenient to follow the path-integral approach of
Refs. \onlinecite{FF,Fidel} and \onlinecite{BOS}.
The free fermionic partition function is
\begin{equation}
Z_0[s]=\int {\cal D}\psi {\cal D}\bar{\psi}~e^{i K_F[\psi] -i
\int d^\nu x j^\mu s_\mu},
\label{z0nud}
\end{equation}
where $K_F$ is a free fermionic action term and 
$j^\mu=\bar{\psi}\gamma^\mu \psi$ ($\nu$ is the dimensionality of
space-time). 

Using gauge invariance (for $\nu=2$ we suppose a gauge invariant
regularization), we have $Z_0[s]=Z_0[s+b]$, where $b$ is a pure
gauge field. Then, up to a global normalization factor, we can write
\begin{equation}
Z_0[s]=\int {\cal D}b|_{\rm pure~gauge} Z_0[s+b].
\end{equation}
This functional integration can also be carried over the whole set of
gauge fields $b$ by imposing an appropiate constraint. For $\nu=2$, the
constraint is given by $\delta[\epsilon^{\mu \nu}\partial_\mu b_\nu]$.
Then,  
exponentiating the delta functional by means of a scalar lagrange
multiplier $\phi(x)$ and shifting $b\rightarrow b-s$, the bosonized
representation is obtained, 
\begin{equation}
Z_0[s]=\int {\cal D}\phi e^{i K_B[\phi] -i \int d^2 x s_\mu
\epsilon^{\mu \nu}\partial_\nu \phi}, 
\label{zfree1d}
\end{equation}
where
\begin{equation}
e^{i K_B[\phi]}=\int {\cal D}b Z_0[b] e^{i \int d^2 x b_\mu
\epsilon^{\mu \nu}\partial_\nu \phi}. 
\label{kb1d}
\end{equation}

For $\nu=3$, the constraint is $\delta[\epsilon^{\mu \nu
\rho}\partial_\nu b_\rho]$  and  the delta functional is exponentiated
by means of a vector field Lagrange mutiplier $A_\mu $. Following the
same steps as before, we obtain  
\begin{equation}
Z_0[s]=\int {\cal D}A e^{i K_B[A] -i \int d^3 x s_\mu \epsilon^{\mu \nu
\rho}\partial_\nu A_{\rho}}, 
\label{zfree2d}
\end{equation}
where
\begin{equation}
e^{i K_B[A]}=\int {\cal D}b Z_0[b] e^{i \int d^3 x b_\mu \epsilon^{\mu
\nu \rho}\partial_\nu A_\rho}. 
\label{kb2d}
\end{equation}

Differentiating $Z_0[s]$, we read from Eqs.\ (\ref{zfree1d}) and
(\ref{zfree2d}) the topological currents that bosonize the fermionic
ones, $\bar{\psi}\gamma^\mu \psi \leftrightarrow \epsilon^{\mu
\nu}\partial_\nu \phi$ and $\bar{\psi}\gamma^\mu \psi \leftrightarrow
\epsilon^{\mu \nu \rho}\partial_\nu A_\rho$, in one and two spatial
dimensions, respectively.

At this point, let us include a general current interaction term,
\begin{equation}
Z[s]=\int {\cal D}\psi {\cal D}\bar{\psi}~e^{i K_F[\psi] + i
I[j^\mu]-i
\int d^\nu x j^\mu s_\mu},
\label{zintnud}
\end{equation}
where $I[j^\mu]$ is represented in terms of a functional Fourier
transform,
\begin{equation}
\exp \left\{ iI[j^\mu] \right\} = {\cal N} \int {\cal D}a_\mu \exp
 \Bigg\{ -i\int d^\nu x h({\bf x}) a_\mu j^\mu  +
iS[a_\mu]
 \Bigg\}.
\label{intnud}
\end{equation}
The constant ${\cal N}$ is chosen such that $I[0]=0$. The function $h({\bf
x})$ is introduced in order to localize the interaction to a spatial region
$\Omega$ (${\bf x}$ is the spatial part of $x$). In other words, $h({\bf
x})$ is a smooth function which is zero, outside $\Omega$, and it grows to
$h({\bf x})=1$, inside $\Omega$. By construction, $I[j^\mu]=0$ for currents
localized outside $\Omega$. The role of the smooth function $h({\bf x})$ is
to implement the adiabatic contact of the interacting region to the
noninteracting one.

Note that for a generic non quadratic interaction $I[j^\mu]$, the
computation of $S(a)$ is, in general, not possible. However, inserting
Eq.~(\ref{intnud}) into Eq.~(\ref{zintnud}) and bosonizing the fermions as
in a free theory with an external source $s_\mu + h({\bf x})a_\mu$ we find 
\begin{equation}
Z[s]=\int {\cal D}\phi~
e^{ iK_B[\phi]+iI[\epsilon^{\mu \nu}\partial_\nu \phi]
-i\int d^2 x  s_\mu \epsilon^{\mu \nu}\partial_\nu \phi},
\end{equation}
for the $1D$ case and 
\begin{equation}
Z[s]=\int {\cal D}A~ 
e^{ iK_B[A]+iI[\epsilon^{\mu \nu \rho}\partial_\nu A_\rho]
-i\int d^3 x  s_\mu \epsilon^{\mu \nu \rho}\partial_\nu A_\rho},
\end{equation}
for the $2D$ case. 
That is, we get the universality of the bosonization rules for the
currents\cite{BOS}.

For $\nu=2$,

\begin{equation}
K_F[\psi]+I[j^\mu]-\int d^2 x  s_\mu j^\mu \leftrightarrow 
K_B[\phi]+I[\epsilon^{\mu \nu}\partial_\nu \phi]
-\int d^2 x  s_\mu \epsilon^{\mu \nu}\partial_\nu \phi.
\label{sbos1d}
\end{equation}

For $\nu=3$,

\begin{equation}
K_F[\psi]+I[j^\mu]-\int d^3 x  s_\mu j^\mu \leftrightarrow 
K_B[A]+I[\epsilon^{\mu \nu \rho}\partial_\nu A_\rho]
-\int d^3 x  s_\mu \epsilon^{\mu \nu \rho}\partial_\nu A_\rho. 
\label{sbos2d}
\end{equation}

At this point some comments are in order. 
Although Eqs. (\ref{sbos1d}) and (\ref{sbos2d}) are similar,
the status of bosonization in $(1+1)$ and $(2+1)$ dimensions is
different. The bosonized action $K[\phi]$ in Eq. (\ref{sbos1d}) is
a local functional of $\phi$ at low as well as at high energies. 
This is a consequence of the constraints imposed by
the space dimensionality, which are not present in the $(2+1)$D case. 
For instance, in a massless $(1+1)$D fermionic free theory it is
possible to build up (zero eigenvalue) bosonic normalizable eigenstates
of the $P_\mu P^\mu$ operator. These modes are represented by the
topologically trivial sector of the corresponding bosonized theory.

In a $(2+1)$D fermionic theory it is not possible to construct such
normalizable eigenstates for any value of the mass. As a consequence, 
one has to be carefull about the meaning of the bosonizing field $A_\mu$.

In Ref. \onlinecite{asymp} we have addressed this question in detail
showing that the topologically trivial sector of the bosonized action
$K_B[A]$ has the vacuum as the only asymptotic state. This can be seen
as follows. 

If a large mass limit is considered, the 
bosonized action for free relativistic fermions takes the form of a
Chern-Simons term and the next correction is a Maxwell term \cite{FF,Fidel}.
Then, we could naively imply the existence of a bosonic mode associated
to the Maxwell-Chern-Simons (MCS) theory. Moreover, other corrections
would be higher derivative terms implying unphysical modes. 
In fact, this is not reliable as all these modes would be at a mass
scale where the approximation is not valid.

In Ref. \onlinecite{BFO} we considered a quadratic approximation instead,
where the full momentum dependence of the fermionic effective action in
Eq. (\ref{kb2d}) were maintained to obtain a nonlocal MCS bosonized
theory. The Schwinger quantization for these kind of theories,
containing nonlocal 
kinetic terms, were developed in Refs. \onlinecite{BOR,AM}. Following these
results, we were able to relate \cite{asymp} the mass weight function of the 
nonlocal MCS bosonized theory and the cross section for fermion pair
creation. Then, as there are no bound states, we were able to imply that
the mass weight function of the bosonized theory has no delta
singularities, i.e., the only asymptotic state in the corresponding
topologically trivial sector is the vacuum. Then, while in $(1+1)$D the
bosonizing field $\phi$ represents some collective excitations of
the fermions, in $(2+1)$D the bosonizing field $A_\mu$ does not.

However, when we study $2$D (parity breaking) fermionic 
systems with gapped excitations, the response to a {\em small} electric
field, or an infinitesimal variation of the chemical potential between
two regions, will be a quasi-equilibrium property dominated by the low
lying energy degrees of freedom. In this regime, the physically
relevant quantities are the currents.

In these cases the bosonization technique will implement a sort of
hidrodynamical approximation where the conserved currents are
represented by $\epsilon^{\mu \nu \rho}\partial_\nu A_\rho$, and the
dynamics is given by Eq. (\ref{sbos2d}). In the last sections we will
follow this route to implement an alternative calculation for the
transverse conductance in $2$D fermionic systems.

%%%%%%%%%%%%%%%%%%%%%%%%%%%%%%%%%%%%%%%%%%%%%%%%%%%%%%%%%%%%%%%%%%%%%%%
\section{Universality of Landauer conductance and bosonization rules
in $1$D systems} 
%%%%%%%%%%%%%%%%%%%%%%%%%%%%%%%%%%%%%%%%%%%%%%%%%%%%%%%%%%%%%%%%%%%%%%%
\label{1D}

Let us summarize in this section the results of Ref. \onlinecite{OMK} where
we have shown that the universality of the free bosonization rules implies
the universality of Landauer conductance at $T=0$, for a general class of
$1D$ systems. There, we have considered a
Fermi-liquid/finite-system/Fermi-liquid structure where the
finite-system presents a quiral symmetry and is 
adiabatically connected to the reservoirs. For a spinless Fermi-liquid, this
amounts to considering for $K_F$, in Eq. (\ref{z0nud}), the usual action for
massless fermions in one dimension, and a smooth $h({\bf x})$ which is zero
outside a finite region extending from $-L/2$ to $+L/2$ (where we have
Fermi-liquid) and it grows to $1$ inside this region (interacting region).

Recalling that the free fermionic effective action in Eq. (\ref{kb1d}) is
quadratic, the path integral over $b_\mu$ can be simply calculated 
to obtain the well known bosonized action
\[
K_B[\phi]=\int d^2 x\, \frac{\pi}{2}\, \partial_\mu \phi \partial^\mu \phi,
\]
and the bosonized equation of motion corresponding to
Eq.~(\ref{sbos1d}) can be written as an anomalous (chiral current)
divergence,
\begin{equation}
\partial_\mu \left[ \partial^\mu \phi+
\frac{1}{\pi}\epsilon^{\mu \nu} \frac{\delta I}{\delta
j^\nu(x)} \right] = - \frac{1}{\pi}E({\bf x},t),
\label{eq:genan}
\end{equation}
where $E({\bf x},t) = \partial_0 s_1({\bf x},t) - \partial_1 s_0({\bf
x},t)$.

Then, following Maslov and Stone's reasoning~\cite{Maslov95}, we
considered an electric field that is switched on until it saturates in
a value $E({\bf x})$. In this case, the large $t$ asymptotic behavior
is given by $\phi({\bf x})=f({\bf x})-k t$. Replacing this behavior in
Eq.\ (\ref{eq:genan}), we have
\begin{equation}
\partial_x \left[\partial_x f +
\frac{1}{\pi} \frac{\delta I}{\delta
j^0(x)} \right] =  \frac{1}{\pi}E({\bf x}).
\label{eq1d}
\end{equation}

Due to causality and the fact that, outside the interacting region,
Eq.\ (\ref{eq:genan}) reduces to the free wave equation, we must have
$f({\bf x})=\pm k {\bf x}+\phi_0 $, the plus (minus) sign corresponding
to the right (left) side of this region. Also notice that the
contribution to the axial current coming from the 
interaction is localized. Therefore, the integration of Eq.\ (\ref{eq1d})
over the spatial coordinate leads to $\partial_x f (b)-\partial_x
f(a)=\frac{1}{\pi}[V(a)-V(b)]$, where $a$ (resp. $b$) lies on the left
(resp. right) side of the finite system, where the electric field is
supposed to be zero. This fixes the constant $k$ to be
$2k=\frac{1}{\pi}[V(a)-V(b)]$ and the electric current (in bosonized
language) results
\be
I=\partial_0 \phi=-k=\frac{1}{2\pi}[V(b)-V(a)].
\label{conductance1d}
\ee
This is the perfect conductance
$e^2/h$ in units where $\hbar = e = 1$. 

If the mass term $m(x)$ is nonzero (a local gap), chiral symmetry is
explicitely broken and the conductance is expected to be suppresed.
This happens in the case of a Peierls dielectric system. 
The {\it quantum} regime of a CDW system is described by a complex
order parameter $\Delta(x)$, representing the lattice degrees of
freedom\cite{Fukuyama85}, whose coupling to the fermions is given by
\begin{equation}
\Delta \bar{\psi}
P_L \psi + \bar{\Delta} \bar{\psi} P_R \psi,
\end{equation}
$P_{R,L} = (1\pm\gamma_5)/2$ are the projectors corresponding to the
right and left modes, respectively.
In this case the fermionic effective action is not known, however, the
bosonized action is known to be a Sine-Gordon model\cite{Col-Man}. For
instance, in a path integral framework, this can be obtained\cite{BQ2},
by taking the Lorentz gauge, and using a lagrange multiplier $\omega$
to write
\begin{equation}
e^{i K_B(\phi)}= \frac{1}{N}\int {\cal D}\psi {\cal
D}\bar{\psi}{\cal D}\omega ~e^{i\int d^2x~\bar{ \psi}i\partial
\!\!\!/\psi +\Delta \bar{\psi}
P_L \psi + \bar{\Delta} \bar{\psi} P_R \psi} 
\delta[\bar{ \psi}\gamma^\mu \psi+\epsilon^{\mu \nu }\partial_\nu \phi -
\partial^\mu \omega].
\label{BQe}
\end{equation}
Studying the behavior of the representation (\ref{BQe}) under
chiral transformations, the well known bosonized action,
\begin{equation}
K_B(\phi)=\frac{\pi}{2} \partial_\mu
\phi \partial^\mu \phi + \frac{A}{2} \left( \bar{\Delta}
e^{-i\beta\phi} +  \Delta e^{i\beta\phi} \right),
\end{equation}
can be derived ($A$ is a renormalization constant). Since in our 
case the order parameter $\Delta$ has
a dynamics given by
\begin{equation}
{\cal L}_{\mbox{\scriptsize{ph}}}[\Delta] = \frac{1}{2v} \left(
\partial_0 \bar{\Delta} \partial_0 \Delta - v^2 \partial_1
\bar{\Delta} \partial_1 \Delta \right) - \frac{\omega_{p}^2}{2v}
\bar{\Delta} \Delta,
\end{equation}
the chiral symmetry is restored.
Then, considering a finite CDW system localized adiabatically
($\Delta(x)\rightarrow h({\bf x})\Delta(x)$) to Fermi-liquid leads,
with an additional (local) current interaction of the form shown in
Eq.~(\ref{intnud}), the field equations of motion lead to an anomalous
chiral current,
\begin{equation}
\partial_\mu j^\mu_{\mbox{\scriptsize{A}}} =  - \frac{1}{\pi}E({\bf
x},t), 
\label{eq:chiralanomaly}
\end{equation}
where the total axial current density components are
\begin{equation}
j^{\mbox{\scriptsize{A}}}_0 = \frac{i\beta h^2}{2v} \left(
\Delta^\dagger \partial_0 \Delta \right) - \left( \Delta \partial_0
\Delta^\dagger \right) + \frac{1}{\sqrt{\pi}}
\frac{\delta I}{\delta j^1(x)} +\partial_0 \phi
\end{equation}
and
\begin{equation}
j^{\mbox{\scriptsize{A}}}_1 = \frac{i\beta v}{2} \left[ \left( h
\Delta^\dagger \right) \partial_1 \left( h\Delta \right) - \partial_1
\left( h \Delta^\dagger \right) \left( h\Delta \right) \right]-
\frac{1}{\sqrt{\pi}} \frac{\delta I}{\delta
j^0(x)}+
\partial_1 \phi.
\end{equation}
In this equation, the chiral current contains the free fermion contribution 
modified by terms coming from the lattice degrees of freedom and the current
interactions, which are localized in the junction. Here again, the
anomalous chiral current divergence leads to a perfect
conductance\cite{OMK}.
For instance, these conclusions hold when forward-scattering
impurities are present in the CDW junction; if impurities are also
present in the Fermi-liquid leads, however, some renormalization of
the conductance is expected, in agreement with the results of
Ref.~\onlinecite{Rejaei96}.

%%%%%%%%%%%%%%%%%%%%%%%%%%%%%%%%%%%%%%%%%%%%%%%%%%%%%%%%%%%%%%%%%%%%%%%%%%%
\section{Universality of transverse conductance and bosonization rules
in $2$D systems }
%%%%%%%%%%%%%%%%%%%%%%%%%%%%%%%%%%%%%%%%%%%%%%%%%%%%%%%%%%%%%%%%%%%%%%%%%%%
\label{2D}

Here, we will show that in the same way that $(1+1)$D bosonization
(cf.\ Eq.\ \ref{sbos1d}) allows a simple calculation of universal
perfect Landauer conductance for $1D$ fermionic systems (cf.\ Eq.\
\ref{conductance1d}), in $(2+1)$D, the bosonized expressions
(\ref{sbos2d}) will allow an exact calculation of universal
transport properties in (parity breaking) fermionic systems with
gapped excitations. In this section we will use, as an example, the
relativistic Dirac field since it is the simplest case that displays the 
required symmetry properties. In the next section we will extend our results
to nonrelativistic systems. 
  
In order to achieve the full bosonization program, we are faced with the
problem of computing the bosonized action $K_B(A)$ for free fermions. This
amounts to computing the functional transverse Fourier transform of the
massive $(2+1)$D fermionic determinant (Eq.\ \ref{kb2d}).
Although a closed expression for $K_B(A)$ is lacking, rather
interesting results have been already established. For instance, in the
infinite mass limit $m\rightarrow \infty $, the effective action
corresponding to the fermionic determinant is given exactly by a local
Chern-Simons term, and its transverse Fourier transform can be computed
straightforwardly, yielding a Chern-Simons bosonized
action\cite{FF}. Another approximation scheme was considered in Ref.\
\onlinecite{BFO}, where the full quadratic part of
the fermionic effective action, corresponding to the exact expression of
the
vacuum polarization tensor, has been taken into account. In this case,
the
approximated bosonized action takes the form of a nonlocal
Maxwell-Chern-Simons term. 
It is worth underlining here that this
approximation has been proven to be very useful in order to discuss on
an equal footing both the massless and the infinite mass limit
corresponding, respectively, to the bosonized actions obtained in
Refs.\onlinecite{Marino} and \onlinecite{FF}. Also, in Ref.
\onlinecite{asymp}, 
we have seen that this nonlocal bosonized theory is physically well defined,
 as the associated mass weight function is positive definite, in
contrast with the result that  would be obtained in any (higher order)
derivative approximation (see also the discussion in section II).

Beyond the quadratic approximation for the fermionic determinant, the
evaluation of the bosonized action is, in general, a difficult task. Using
the results of Ref.\onlinecite{Silvio} , some general features of the
bosonized action can be obtained. Recently, we have shown\cite{BOS} that
the bosonized action $K_B(A)$ in Eq. (\ref{sbos2d}) can be cast in
the form of a pure Chern-Simons term, up to a nonlinear and nonlocal
redefinition of the gauge field. In this way we can separate the topological
information from the particular (and unknown) details of the bosonized
action. For our present purposes this representation is not needed. However,
we will take advantage of the general structure underlying this separation.

Now, we will present a relationship between the result of Ref.
\onlinecite{BOS}, i.e., the universal character of the current
bosonization rules for 2D interacting fermionic systems (cf. Eq.\
(\ref{sbos2d})), and the universality of transverse conductance between
``perfect Hall regions'' (where the parity breaking parameter
goes to infinity). This is the $2$D counterpart of the relationship 
between universality of current bosonization rules and Landauer
conductance, in $1$D systems.

At this point one question naturally arises: how can we try to obtain some
exact result if we do not even know the exact expression for the bosonized
kinetic action $K_B[A]$ ? The key point is that the unknown terms in
the bosonized kinetic action 
will have the same form of the bosonized interaction term, and when
looking at universal behavior, 
the detailed form of these terms will be irrelevant. Physically, in
analogy with the $1D$ case, 
where the Fermi-liquid in the reservoirs imposes strong constraints on the
Landauer conductance, we will see that perfect Hall regions will
impose strong constraints on the transverse conductance.

Let us consider relativistic $2D$ fermions with a position
dependent (positive definite) mass $m({\bf x})$, 
\begin{equation} 
K_F=\int d^3x~\bar{\psi}(i\partial
\!\!\!/+m({\bf x}))\psi. \label{kfinhom} 
\end{equation}
We will suppose that there are at least two disconnected regions or
``islands'' where the gap $m({\bf x})$ goes to infinity.  The multiply
connected region around the islands 
will be called $\Omega$. For definitness, we will consider a mass parameter
which takes a value $M$ ($M\rightarrow \infty$), inside the islands, while it
rapidly decreases to a finite value $m$ outside them.

Firstly, since the bosonized action $K_B(A)$ is obtained from a functional
transverse Fourier transform of the fermionic determinant (cf.\ Eq.\
(\ref{kb2d})), it is gauge invariant. Secondly, it is easy to see that in the
case of uniform mass, the bosonized kinetic term $K_B(A)$ can be written as
(see for example Refs.\ \onlinecite{BOS,Silvio})
\begin{equation}
K_{\rm hom}[A]= \frac{1}{\eta} S_{CS} + \tilde{R}_{\rm hom} 
[\epsilon \partial A],
\end{equation}
where
\be
S_{CS}=\frac{1}{2}\int d^3 x A_\mu\epsilon^{\mu\nu\rho} \partial_\nu A_\rho,
\ee
$\eta= \frac{M}{|M|}\frac{1}{4\pi}$
and $\tilde{R}_{\rm hom}$ goes to zero as the mass goes to infinity. A
similar conclusion applies to the case where the mass is $x$-dependent, i.e.,
when $m({\bf x})$ is replaced by $\lambda m({\bf x})$, and
$\lambda \rightarrow \infty$, we have (including a gauge fixing factor in Eq.\ 
(\ref{kb2d}))
\begin{equation}
\lim_{\lambda\rightarrow \infty } \exp{i K_B(A)}=\int {\cal D}b_\mu
F(\partial 
b)~e^{i \eta S_{CS}(b)+i\int d^3x~\varepsilon ^{\mu \nu \rho }A_\mu
\partial_\nu b_\rho },  \label{lef}
\end{equation}
where we have used that the large distance behavior of the fermionic
effective action is dominated by $\eta S_{\rm CS}(b)$. Integrating over
$b_\mu$ we get,
\begin{equation}
\lim_{\lambda\rightarrow \infty} K_B(A)=\frac{1}{\eta} S_{CS}(A)\;.
\label{limfree}
\end{equation}
Thus,  the bosonized action contains a local Chern-Simons term,
corresponding to the bosonization in the infinite mass limit, first
obtained in Ref.\ \onlinecite{FF}. Therefore, based on gauge invariance
of the bosonized action, for any finite value of $\lambda$ we can write
\begin{equation}
K_B(A)=\frac 1\eta S_{CS}(A)+R[\epsilon \partial A]\;,
\label{kbinhom}
\end{equation}
where 
\begin{equation}
\lim_{\lambda \rightarrow \infty }R[\epsilon \partial A]=0\;.
\end{equation}
In particular, setting $\lambda=1$ in Eq. (\ref{kbinhom}), 
a pure local Chern-Simons term, with parameter $\frac 1\eta$, can be
isolated from the bosonized kinetic action for fermions with mass
parameter $m({\bf x})$.
The remaining part is a gauge invariant
functional $R[\epsilon \partial A]$ where every term contains a
nontrivial dependence on $m({\bf x})$, which goes to zero when 
$m({\bf x})$ is replaced by $\lambda m({\bf x})$, and $\lambda
\rightarrow \infty$. Then, we expect that when considering any local
derivative expansion of $R$, inside the islands, where the mass
parameter takes the value $M$ ($M\rightarrow \infty$), the bosonized
kinetic action takes the form of a pure Chern-Simons term with
parameter $\frac 1\eta$. In other words, the functional $R[\epsilon
\partial A]$ is localized in $\Omega$, i.e., when the support of 
$\epsilon \partial A$ is localized outside, we have $R[\epsilon
\partial A]=0$. Equivalently, if we write
\begin{equation}
R[\epsilon \partial A]=\int d^3x {\cal R} (\epsilon \partial A),
\end{equation}
the local density ${\cal R}$ is zero when evaluating $\epsilon \partial
A$ in a point outside $\Omega$.

In order to see this behavior more clearly, let us take the
representation
(\ref{sbos2d}), replace the expression (\ref{kfinhom}) of the fermionic
partition function and integrate over $b_\mu$,
\begin{equation}
e^{i K_B(A)}=\int {\cal D}b_\mu F(\partial b)~\int {\cal D}\psi {\cal
D}\bar{\psi}~e^{i\int d^3x~\bar{ \psi}(i \partial
\!\!\!/+m(x,y)+ib\!\!\!/)\psi } e^{i\int
d^3x~\varepsilon ^{\mu \nu \rho }A_\mu \partial _\nu b_\rho }.
\end{equation}
Proceeding in a similar way to Burgess and Quevedo\cite{BQ2}, we can take
the Lorentz gauge, and use a lagrange multiplier $\omega$ to express
\begin{equation}
e^{i K_B(A)}= \frac{1}{N}\int {\cal D}\psi {\cal
D}\bar{\psi}{\cal D}\omega ~e^{i \int d^3x~\bar{ \psi}(i \partial
\!\!\!/+m({\bf x}))\psi } 
\delta[\bar{ \psi}\gamma^\mu \psi+\epsilon^{\mu \nu \rho} \partial_\nu 
A_\rho - \partial^\mu \omega],
\label{alter}
\end{equation}
where $N$ is chosen such that $K_B(0)=0$,
\begin{equation}
N=\int {\cal D}\psi {\cal
D}\bar{\psi}{\cal D}\omega ~e^{i\int d^3x~\bar{ \psi}(i\partial
\!\!\!/+m(x,y))\psi }
\delta[\bar{ \psi}\gamma \psi -\partial \omega].
\end{equation}
(Eq. (\ref{alter}) should be compared with its $1D$ counterpart
(\ref{BQe})). 

Using a lattice regularization of the path integral in Eq.\ (\ref{alter}), we
see that when $\epsilon \partial A$ is localized outside $\Omega$, the
contribution to the nontrivial $A$ dependence comes from those $\psi$'s which
are coupled to $m({\bf x})$ outside $\Omega$. Therefore, for 
$\epsilon \partial A$ localized on this region (inside the islands), we can
compute (\ref{alter}) replacing $m({\bf x})$ by $M$ ($M\rightarrow \infty$),
and a Pure Chern-Simons action is obtained. Summarizing, because of
locality, the bosonized action is basically a pure
Chern-Simons term on the islands, the only possible excitations there
corresponding to (nondissipative) currents which are transverse to the
external electric field.

Including a current dependent interaction $I[j]$, and using the universality
of the bosonization rules for the currents (cf. Eq.(\ref{sbos2d})), we are
left with the complete bosonized action for a 2D (relativistic) fermionic
system
\begin{equation}
\frac 1\eta S_{CS}(A)+R[\epsilon \partial A]+I[\varepsilon \partial A]+
\int d^3x~s^\mu \varepsilon _{\mu \nu \rho }\partial ^\nu A^\rho, 
\label{o1}
\end{equation}
where $s_\mu$ is the external source (we will suppose that the external
electric and magnetic fields are localized in $\Omega$). Note that the
unknown terms in the bosonized action have the same form of the bosonized
interaction term, and they have the same localization properties. As
anticipated, this is the reason why we can obtain {\it exact} results, when
looking for universal behaviors.

The corresponding equations of motion are
\begin{equation}
\frac{1}{\eta}\epsilon^{\mu \nu \rho} 
\partial_{\nu} A_\rho - \epsilon^{\mu \nu \rho} 
\partial_{\nu} \frac{\delta (R+I)}{\delta j^{\rho}(x)}
=-\epsilon^{\mu \nu \rho} \partial_{\nu} s_\rho.
\label{mov}
\end{equation}
(here, the combination $\epsilon^{\mu \nu \rho} \partial_{\nu}
A_\rho$ has been called $j^{\mu}$).
Taking the $i$ component of Eq.\  (\ref{mov}), and considering
stationary external sources, we are left with the equation
\begin{equation}
\frac{1}{\eta}\partial_k A_0 -
\partial_k \frac{\delta (R+I)}{\delta j^0(x)}=-\partial_k s_0,
\end{equation}
and integrating on a curve that goes from a point ${\bf a}$ in the
interior of a perfect Hall region (island) to a point ${\bf b}$ on another,
\begin{equation}
\frac{1}{\eta}\left (\int d{\bf x} . {\bf \nabla} A_0 \right)-
\left. \frac{\delta (R+I)}{\delta j^0(x)}\right|_{{\bf a}}^{\bf b}=
V({\bf b})-V({\bf a}). 
\end{equation}
The first term corresponds to the bosonized expression of the
transverse current,
\begin{equation}
I_t=\int d{\bf n} \cdot {\bf j}, 
\end{equation}
where {d\bf n} is a normal element with respect to the integration
curve,
whose components are $d n_i=\epsilon_{ij} dx_j$.
On the other hand, the last term of the first member is zero as
the interaction ($I$) and the bosonized kinetic part which is not
Chern-Simons ($R$) are localized in $\Omega$. The second member is the
electric potential difference between both regions.

Summarizing, our result is that the transverse current $I_t$, between
two ``perfect Hall regions'' does not depend on the current interactions
localized outside them, nor on the particular geometry of these
regions, and is given by
\begin{equation}
I_t=\frac{1}{4\pi} \left(V(b)-V(a)\right),
\end{equation}
which corresponds to a transverse conductance
$\frac{1}{2}\left(\frac{e^2}{h}\right)$ (in ordinary units).
As before, the fundamental role played by the adiabatic
transition between the interacting and the noniteracting regions
becomes evident. It permits a unified treatment of the perfect Hall
and interacting regions in a single theory, establishing a particular
matching between them, which corresponds to nondissipative contacts.

The particular value of the conductance $(1/2)e^2/h$ is entirelly due
to relativistic invariance. Moreover, to the best of our knowledge, the
massive Dirac fermion is the only local system with transverse
conductance $1/2$, and {\em finite} (although not universal)
longitudinal conductance. For this reason, it could be related
to models describing quantum critical properties for transitions
between plateaux's in the Quantum Hall Effect\cite{duality}.
  
Now, let us suppose the case where $\Omega$ is a circle.
Taking  $\mu=0$ in Eq.\ (\ref{mov}), and integrating over a region $S$
containing this circle,
\begin{equation}
\frac{1}{\eta}\int d^2x \epsilon^{ik} 
\partial_{i} A_k - \epsilon^{ik} 
\partial_{i} \frac{\delta (R+I)}{\delta j^k(x)})
=-\int d^2x \epsilon^{ik} \partial_{i} s_k. 
\end{equation}
Using Stoke's theorem,  we can pass to an integration over the border
of $S$, which 
is contained outside $\Omega$. There, the local densities ${\cal R}$ and 
${\cal I}$ are zero obtaining
\begin{equation}
\frac{1}{\eta}\int d^2x \epsilon^{ik} 
\partial_{i} A_k=-\frac{1}{4\pi}\oint ds \epsilon^{ik} \partial_{i} s_k. 
\end{equation}
The first member is the bosonized expression for the electric charge
contained in $S$, while the second member is the magnetic flux
through $S$. In the noninteracting case, this corresponds to the
Aharonov-Casher result\cite{Aharonov-Casher} for the ground state of a
relativistic 
fermionic system, and comes from the spectral asymmetry associated to
fermions in the presence of an external ({\bf x}-dependent) magnetic
field. With this calculation  we are showing that this relationship is
an exact and universal result, independent
of the current interactions in $\Omega$.

%%%%%%%%%%%%%%%%%%%%%%%%%%%%%%%%%%%%%%%%%%%%%%%%%%%%%%%%%%%%%%%%%%%
\section{Nonrelativistic fermions in a magnetic field}
%%%%%%%%%%%%%%%%%%%%%%%%%%%%%%%%%%%%%%%%%%%%%%%%%%%%%%%%%%%%%%%%%%%
\label{nonrel}

In the previous section we have called ``perfect Hall regions'' those regions
where the (parity breaking) mass parameter goes to infinity. In these regions,
the bosonized action is a pure Chern-Simons term. We have found that 
the conductance between these regions is $\frac{1}{2}\frac{e^2}{h}$,
whatever the form of the current interactions considered.
This value of the conductance is a characteristic of relativistic
fermions in vacuum. However, in order to make contact with the quantum Hall
effect (integer or fractional) we should consider nonrelativistic fermions
at finite density submetted to a magnetic field, perpendicular to the plane.
Here, in a similar way to the relativistic case, we shall consider arbitrary 
interactions localized on a region $\Omega$ that is adiabatically connected to 
regions (``islands'') where the system displays an exact
integer Landau quantization. To implement this model, let us consider the
following action
\be
S=\int d^2x dt~\psi^*(x) \left\{ -i\partial_t + e s_0
+\mu+ \frac{1}{2m} \left[-i\vec\nabla+e(\vec d+\vec{ s})\right]^2 \right\}
\psi(x) + I(\psi^*\psi),
\ee
where $\vec\nabla\times\vec d=B_{\rm ext}({\bf x})$, $I(\psi^*\psi)$ is an 
arbitrary nonrelativistic interaction localized in $\Omega$ (possibly
nonlocal and non quadratic), and $s_\mu$ is a source introduced to
prove the system. In $\Omega$, we shall consider a position dependent magnetic 
field $B_{\rm ext}({\bf x})$, 
which changes adiabatically to some constant value on the islands, where
the chemical potential is adjusted in order to have the first Landau
level completelly filled. Note that outside the islands, in general,
there is no Landau quantization. 

In Ref. \onlinecite{Linhares} we have shown that for the nonrelativistic 
interactions consider in this model the current  
bosonization rules,
\bea
\rho(x) &\rightarrow& \vec\nabla\times \vec A \\
j_i(x) &\rightarrow& \epsilon_{ij}E_j(A),
\eea
are universal, and the bosonized action can be cast in terms of a gauge
field $A_\mu$ in the following form
\be
S_{\rm bos}=K_B[A]+ I(\vec\nabla\times\vec A), 
\ee 
where 
\be e^{ i K_B(A)}=\int {\cal D}b_\mu 
e^{{\rm Tr}\ln\left(
-i\partial_t+e b_0 + \mu+ \frac{1}{2m}
\left[-i\vec\nabla+e(\vec d+\vec{ b})\right]^2\right)+i\int d^2x d\tau
b_\mu\epsilon_{\mu\nu\rho}\partial_\nu A_\rho }.
\label{lmn}
\ee

The nonrelativistic fermionic determinant is a very complicated object and
no exact analytic result is known for the general case. In the gaussian
approximation, when
$B_{ext}\approx 0$, the spectrum is gapless and
there is no signal of topology in the structure of the
determinant\cite{Linhares}. However when $B_{\rm ext}$ is large and
varies slowly, the situation is completelly different since the Landau
quantization opens gaps in the spectrum. In this case, it is possible
to make a gradient expansion of the determinant
obtaining\cite{determinant1,determinant2,LopezFradkin}
\bea
\lefteqn{
{\rm Tr}\ln\left(
-i\partial_t+e b_0+\mu+ \frac{1}{2m}
\left[-i\vec\nabla+e(\vec d+\vec{ b})\right]^2
\right)=-i \int d{\bf x} dt } \nonumber \\ 
&&\left\{
\frac{e^2}{4\pi}\gamma b_\mu\epsilon_{\mu\nu\rho}\partial_\nu b_\rho - 
\frac{e^2}{2\pi m}(\frac{\gamma^2}{2}-\gamma) 
(\vec\nabla\times\vec b+B_{\rm ext})^2
+\frac{e^2}{2\pi}\gamma b_0 B_{\rm ext}
\right\}+O(|\frac{eB_{\rm ext}}{m}|^{-1}), 
\label{trln}
\eea
where
\be
\gamma=-\sum_{n=0}^\infty \Theta
\left[\mu+e b_0-\left(n+\frac{1}{2}\right)\omega_c\right],
\ee
and $\omega_c=e B_{\rm ext}/m$. 

Due to the presence of the function $\gamma$, Eq.\ (\ref{trln}) is an
extremely complicated non quadratic functional of the field. However, in the
limit of constant (and large) $B_{\rm ext}$, we can adjust the chemical
potential in such a way that $\gamma=-1$. This procedure is equivalent to
projecting the effective action onto the first Landau level. In this
approximation, it is simple to integrate the field $b_\mu$ (upon gauge
fixing) obtaining the bosonized action
\be
\lim_{B_{\rm ext}/m>>1}K_B(A)\equiv K_\infty(A)=\left(\frac{2\pi}{e^2}\right)
\int d^2x d\tau~  
\left\{
\frac{1}{2} A_\mu\epsilon_{\mu\nu\rho}\partial_\nu A_\rho - 
\frac{3}{2m}(\vec\nabla\times\vec A)^2
+ A_0 B_{\rm ext}\right\}.
\ee
Note that this action is not of the pure Chern-Simons
form. This is related to the possibility of inducing currents by means of 
magnetic field inhomogeneities. Also, the last term ($A_0B_{\rm ext}$)
indicates that we are considering fermions at finite density, where the
ground state charge density is proportional to $B_{\rm ext}$. However, the
main point at this moment is that, since the exact bosonized action $K_B$ 
is gauge invariant 
(cf.\ Eq.\ (\ref{lmn})), it can be cast in the form
\be
K_B(A)=K_\infty(A)+R(\vec{j}(A), \rho(A)), 
\ee
where
\be
\lim_{eB_{\rm ext}/m\rightarrow\infty}R(\vec{j}(A), \rho(A))=0. 
\ee
(here we have written the gauge invariant variables, in the functional $R$, 
in terms of $\rho(A)$ and $\vec{j}(A)$, the bosonized density and currents).
It is also important to notice that while Lorentz covariance is lost 
(since the system is nonrelativistic), the bosonized action remains
gauge invariant. Gauge symmetry is precisely one of the ingredients we
need to show universal behavior in $2$D systems.

Including the nonrelativistic fermionic interactions and using the
universality of the bosonization rules for the currents, we can write the
complete bosonized action for $2$D nonrelativistic fermions as
\be
S_{\rm bos}(A)=K_\infty(A)+ R[\vec{j}(A), \rho(A)] + I[\rho(A)]+i\int d^3x
s^\mu \epsilon_{\mu\nu\rho}\partial^\nu A^\rho. 
\label{bosnonrel}
\ee

For time-independent external sources, the stationary equations 
of motion corresponding to the bosonized action (\ref{bosnonrel}) read
\bea
\frac{\delta S_{\rm bos}}{\delta A_0}=0
&\rightarrow&
\frac{2\pi}{e^2}\left[\vec\nabla\times \vec A+B_{\rm ext}\right]+ 
\vec\nabla\times\frac{\delta R}{\delta \vec{j}}=
-\vec\nabla\times\vec s,  
\label{eq1} \\
\frac{\delta S_{\rm bos}}{\delta A_k}=0
&\rightarrow& 
\frac{2\pi}{e^2}\left[\vec\nabla A_0
-\frac{3}{m}\vec\nabla(\vec\nabla\times A)\right]-
\vec\nabla\left(\frac{\delta (R+I)}{\delta \rho}\right)=\vec\nabla s_0. 
\label{eq2}
\eea

Replacing (\ref{eq1}) in (\ref{eq2}) we find
\be
\frac{2\pi}{e^2}\vec\nabla A_0+
\frac{3}{m}\vec\nabla\left(\vec\nabla\times\frac{\delta R}{\delta \vec{j}}
+\frac{2\pi}{e^2}B_{\rm ext}\right)-
\vec\nabla\left(\frac{\delta (R+I)}{\delta \rho}\right)=\vec\nabla\left( s_0-
\frac{3}{m}\vec\nabla\times\vec s\right). 
\label{eqnonrel}
\ee
The last term ($\vec\nabla\times \vec s$) comes from the fact that in this
system it is possible to induce currents by applying an inhomogeneous magnetic
field perpendicular to the plane. To calculate the conductance we
consider only an external electric field ($\vec\nabla\times \vec
s=0$). Then, integrating the last equation along a line with endpoints
on different islands, where the Landau quantization is exact, we obtain
\be
\frac{2\pi}{e^2}\int d\vec x\cdot\vec\nabla A_0+ \left.
\frac{3}{2m}\left(\vec\nabla\times\frac{\delta R}{\delta \vec{j}}
+B_{\rm ext}\right)\right|_{\bf a}^{\bf b}-
\left.\frac{\delta (R+I)}{\delta \rho}\right|_{\bf a}^{\bf b}=V({\bf b})-V({\bf
a}). 
\ee
The first term is the bosonic version of the transverse current, the second
term is zero since, on the islands, the local density associated to $R$
is zero ($R$ is localized in $\Omega$) and $B_{\rm ext}$ is a constant there.
The third term is also zero since the interactions are also localized
in $\Omega$. So, turning back to usual units we find
\be
I_t=\frac{e^2}{h}\left\{V({\bf b})-V({\bf a})\right\}. 
\ee
This means that the transverse conductance is exact and universal (and of
course has the correct coefficient $\frac{e^2}{h}$). 

Although in this
example we have evaluated the conductance between regions in the integer
quantum Hall state, a straightforward generalization to the
fractional Quantum Hall effect (Laughlin or Jain states) can be done by
using the fermionic 
action proposed by Ana Lopez and Eduardo Fradkin in Ref.\
\onlinecite{LopezFradkin}. In this case, we have to 
deal with an extra Chern-Simons gauge field (called statistical field)
that esentially works attaching  fluxes to the charges, building up
in this way 
the concept of composite fermions\cite{Jain}.
The main idea is that, in the mean field approximation, 
the ``ficticious magnetic field'' produced by the statistical
Chern-Simons field spreads out, and combines with the real external
field to produce an effective magnetic field given by
\begin{equation}
B_{\rm eff}= B+\langle {\cal B} \rangle= B-2\pi (2 s)\bar\rho,
\label{Beff}
\end{equation}   
where $\langle {\cal B} \rangle=-2\pi (2s)\bar\rho$ is the mean value
of the statistical  
magnetic field, in the mean field approximation ($\bar\rho$ is the mean
density  and $2 s$ 
counts the number of elementary quantum fluxes attached to each particle). 

Then, the system displays an effective integer Landau level quantization 
with an effective magnetic field $B_{\rm eff}$ given by Eq. (\ref{Beff}). 
Thus, having  Landau gaps, it is possible to develope a gradient
expansion to evaluate  
the fermionic determinant. 
The result is almost the same of Eq. (\ref{trln}), so we  can  project
the system into  
the ``first'' Landau level and adjust the chemical potential to have an
effective filling factor $\nu_{\rm eff}=1$.  
The main difference with the integer case is that when 
expressing  the filling factor in terms of the original ``real''
magnetic field, one has  
$\nu=\frac{1}{2 s+ 1 }$, this corresponds to the main sequence of plateaux's 
described by the Laughlin wave functions. From this point of view, the
mean field composite fermion  
theory for the fractional 
quantum Hall effect is essentially the theory of the integer effect
with a renormalized magnetic field.     
 
This model for the fractional QHE can be bosonized in exactly the same
way as in the integer 
case. Now, the equation of motion, derived from the bosonized action, leads to
the transverse current
\be
I_t=\left(\frac{1}{2 s+1}\right)\left(\frac{e^2}{h}\right)\left\{V({\bf
b})-V({\bf 
a})\right\}.
\ee
Thus, with the simple argument of universality of the bosonization rules in
$2$D and with the assumption that regions with perfect integer or fractional 
Hall quantization (used to measure the conductance) are adiabatically
connected to a non quantized interacting region, we were able to deduce that 
the transverse conductance is exact and universal.

%%%%%%%%%%%%%%%%%%%%%%%%%%%%%%%%%%%%%%%%%%%%%%%%%%%%%%%%%%%%%%%%%%%%%
\section{Summary and Conclusions}
%%%%%%%%%%%%%%%%%%%%%%%%%%%%%%%%%%%%%%%%%%%%%%%%%%%%%%%%%%%%%%%%%%%%%
\label{conclusion}

In this paper we have presented a simple way to study
universal transport properties for some bidimensional systems, 
relying on recent studies on the bosonization program.

In one dimension, bosonization is a natural way to display the anomalous 
properties of chiral symmetry, 
the physical principle behind the universal Landauer
conductance in Fermi-liquid/finite-system/Fermi-liquid structures.
In two dimensions, bosonization is a convenient way to display
the parity breaking properties of the system and the underlying 
gauge symmetry, that is, the physical principle behind the universal
quantization of the Hall conductance in the presence of impurities.

Although a complete bosonization in $2$D is not yet available, we
were able to show exact and universal quantization of the transverse 
conductance between perfect Hall regions, for a whole class 
of current interactions. 
In this regard, we note that the unknown functional $R$ in the bosonized 
kinetic action 
(cf. Eqs. (\ref{o1}) and (\ref{bosnonrel})), has the same form of the 
localized 
interaction term. Then, when looking at universal behavior, the exact form of this 
unknown term is irrelevant, all we need is that this term be 
localized outside the perfect Hall regions.

We can summarize our results by saying that, in the same way that in 1D
systems, the Fermi-liquid in the reservoirs imposes strong constraints 
on Landauer's conductance, in $2$D systems, the perfect Hall regions
impose strong constraints on the transverse conductance.
We have also seen that for this properties be operating, 
the transition between interacting and noninteracting regions should be
adiabatic, this corresponds to nondissipative contacts.
As a by product we also showed that the Aharonov-Casher relation
between charge and magnetic flux, originally deduced for free
relativistic fermions in a magnetic field\cite{Aharonov-Casher}, are
universal.

In the case of relativistic fermions, the particular parity breaking 
properties of the 
ground state leads to a (half) perfect transverse conductance. 

In the nonrelativistic case, we showed the universality of the 
integer and fractional 
Hall conductance. This fact clearly shows that relativistic covariance 
is not important to deduce universal transport. 
Actually, these properties have to do with general structures, 
such as universality of the topological 
bosonized currents, localization and gauge invariance properties 
of the bosonized action, 
the presence of parity breaking topological terms and the presence of a
gap in the charged degrees of freedom. 

Since these results are independent of the shape of the regions used to
measure the conductance, we could consider, for instance, two regions
in a perfect quantum Hall state, adiabatically connected by a 
straight line potential barrier, where impurities and phonon
interactions are present.  
This tunnel junctions
are in fact experimentally available \cite{stormer}. As long as
the barrier interaction can be 
considered adiabatically switched off on the bulk, the Hall conductance 
between the QHE regions should be exact and universal, since it is
dominated by the bulk states.

In Refs.\
\onlinecite{Luca1,Luca2,Luca3} we have also obtained this universality using an
exact model calculation. In those references, a one dimensional effective
field theory was used to evaluate transport properties for a barrier
between quantum Hall samples, relating the exact quantization 
to the chiral properties of the model. In the present work we showed an
alternative derivation, using general assumptions and no
model calculations, extending these properties to the fractional QHE
case, for example.

Finally we would like to point out that, although the bosonization program 
in higher dimensions is not fully developed, it is an extremelly
usefull technique to  obtain universal properties of strongly
correlated fermions in a simple and transparent way.

%%%%%%%%%%%%%%%%%%%%%%%%%%%%%%%%%%%%%%%%%%%%%%%%%%%%%%%%%%%%%%%%%%%%%%%
\acknowledgments
We are indebted to Eduardo Fradkin for very useful comments about the
manuscript. We also thank Cesar Fosco for fruitfull discussions.

D.G.B. is partially suported by NSF, grant number NSF DMR98-17941 
at UIUC, by the Brazilian Agency CNPq through a post-doctoral
fellowship and by the University of the State of Rio de Janeiro, RJ, 
Brazil.

%%%%%%%%%%%%%%%%%%%%%%%%%%%%%%%%%%%%%%%%%%%%%%%%%%%%%%%%%%%%%%%%%%%%%%%

\end{document}